\documentclass[preprint,amsmath,aps,prstab,longbibliography]{revtex4-1}
\usepackage{graphicx} 
\usepackage{epstopdf}

\begin{document}

\title{Application of transfer matrix and transfer function analysis to grating-type dielectric laser accelerators: ponderomotive focusing of electrons}

\author{Andrzej Szczepkowicz}

\affiliation{Institute of Experimental Physics, University of Wroclaw, Plac Maksa Borna 9, 50-204 Wroclaw, Poland}

\date{\today}

\begin{abstract}
The question of suitability of transfer matrix description of electrons traversing grating-type dielectric laser acceleration (DLA) structures is addressed. It is shown that although matrix considerations lead to interesting insights, the basic transfer properties of DLA cells cannot be described by a matrix. A more general notion of a transfer function is shown to be a simple and useful tool for formulating problems of particle dynamics in DLA. As an example, a focusing structure is proposed which works simultaneously for all electron phases. 
\end{abstract}

\maketitle

\section{Introduction}\label{sect-intro}

Several recent proof-of-principle experiments demonstrate the possibility of accelerating electrons
in a laser-driven dielectric structure \cite{dla2014,dla2016}. One class of such dielectric
laser accelerator (DLA) structures is the grating-type structure, in which a unit cell is iterated in one dimension,
as in the recently developed single grating, dual-grating, and dual pillar structures \cite{dla2016}.
On basis of these successful experiments, compact laser driven accelerators are envisioned
(see for example Fig.~4.\ in Ref.\ \cite{BreuerHommelhoff2013}). A working device will require,
in addition to acceleration, beam focusing, and possibly beam diagnostics sections and feedback beam steering.
To design a complete DLA beamline, a mathematical description of electron trajectory throughout the whole device
is necessary. For conventional radio-frequency (RF) accelerators, several mathematical tools were developed over 
the years to effectively describe the single particle and beam trajectories \cite{BrownServranckx1984,Wille2000,Wiedemann2015,uspas}. One such tool is the transfer matrix; it used to describe the particle transfer properties of the various building blocks of a beamline. For grating-type DLAs, the natural building block is the unit cell of the grating \cite{JoannopoulosJohnson2008}. Here, interesting questions arise: what are the particle transfer
properties of a DLA unit cell, and can they be described by a matrix? This problem has been partially addressed in Ken Soong's PhD thesis \cite{SoongPhD2014}, where the transfer matrix of a unit cell of a double-grating accelerator structure is calculated.
In this pioneering work the adequacy of linear approximation is not discussed, and a 25-attosecond electron bunch is assumed, with length less than 1\% of the grating period, evading the problem of distribution of phases. 
The purpose of the present work is to pursue further this interesting idea.

\section{The transfer matrix formalism}\label{sect-tm}

In conventional RF accelerators particle motion is described
relative to a \emph{reference trajectory} \cite{BrownServranckx1984,HemsingStupakov2014}.
The reference trajectory defines a coordinate system which is in general curvilinear,
with the distance along the trajectory described by coordinate $S$ (following the notation in Ref.\ \cite{HemsingStupakov2014}), and with orthogonal coordinates $x,y$ describing the particle position in the transverse plane.
The particle on the reference trajectory has \emph{reference energy} ${\cal E}_0$ 
(corresponding to \emph{reference momentum} $p_0$). The relative position of electrons on
the reference trajectory with respect to the beam center is measured by $s$.
The electron location in the six-dimensional phase space comoving with the electron beam
is characterized by the vector $\vec X=(x,x',y,y',s,\eta)^{\mathrm T}$ \cite{HemsingStupakov2014},
where $x'=dx/dS$ and $y'=dy/dS$ are the small angles of deflection from the reference trajectory,
and $\eta=\Delta{\cal E}/{\cal E}_0$ is the relative energy deviation
(other authors \cite{BrownServranckx1984,Wille2000} use relative momentum deviation $\delta=\Delta p/p_0$ instead of $\eta$;
in the ultrarelativistic limit $\eta=\delta$).
Note that all coordinates of $\vec X$ are small and the \emph{reference particle} is described by
$\vec X=(0,0,0,0,0,0)^{\mathrm T}$. 

In conventional RF accelerators, basic properties of a beamline section can be described by
first-order beam transport optics \cite{BrownServranckx1984}, using linear approximation:
\begin{equation}
\vec X_2={\cal R}\vec X_1,
\end{equation}
where $\vec X_1$ describes a particle at the entrance of the section,
$\vec X_2$---at the exit of the section, and $\cal R$ is a linear transfer function, 
which is represented by a $6\times6$ \emph{transfer matrix}.
Phenomena not captured by this approximation can be described by second-order optics
\cite{BrownServranckx1984} or by detailed numerical particle tracing.

\begin{figure}
\centering
\includegraphics[scale=0.7]{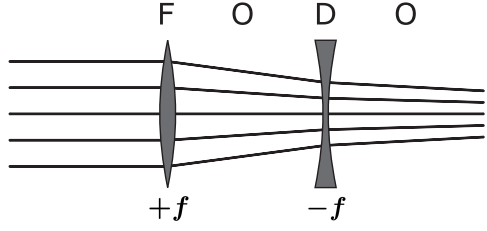}%
\caption%
{%
 \label{fig-fodo}%
 The principle of FODO focusing. The best known realization of this principle is the alternating gradient focusing
 used in conventional RF accelerators.
}
\end{figure}

Often in the literature a reduced form of the $\cal R$ matrix is used \cite{BrownServranckx1984,Wille2000,Wiedemann2015,uspas},
where, as a starting point of the analysis, chromatic effects are neglected ($\delta=0$, $\eta=0$),
and only $(x,x')$ phase plane is considered:
\begin{equation}
\left(\begin{array}{c}x_2\\x'_2\end{array}\right)=
\left(\begin{array}{cc}R_{11}&R_{12}\\R_{21}&R_{22}\end{array}\right)
\left(\begin{array}{c}x_1\\x'_1\end{array}\right).
\end{equation}
In the context of classical optics, such formulation is called ray transfer matrix analysis (or ABCD matrix analysis)
and is used to describe the propagation of light rays and Gaussian beams in the paraxial approximation
\cite{Brooker2007}. Beam transfer through a thin lens of focal length $f$ is described by the matrix
\begin{equation}
\label{eq-matrixF}
{\cal F}=\left(\begin{array}{cc}1&0\\-1/f&1\end{array}\right)
\end{equation}
($1/f$ is called \emph{optical power} or \emph{focusing power}).
A free drift region of length $s$ with no optical elements is described by the transfer matrix
\begin{equation}\label{eq-matrixO}
{\cal O}=\left(\begin{array}{cc}1&s\\0&1\end{array}\right).
\end{equation}
One of the common building blocks used in design of RF accelerator beamlines is the FODO array
(focusing--drift--defocusing--drift) \cite{BrownServranckx1984,Wille2000,Wiedemann2015,uspas},
which has an overall focusing effect, see Fig.~\ref{fig-fodo}. 
The transfer function of the FODO array is the mathematical composition of 
the transfer functions of its four building blocks. Composition of linear functions
is equivalent to matrix multiplication:
\begin{eqnarray}
{\cal R}_\mathrm{FODO}&=&{\cal ODOF}=
\left(\begin{array}{cc}1&s\\0&1\end{array}\right)
\left(\begin{array}{cc}1&0\\-1/(-f)&1\end{array}\right)
\left(\begin{array}{cc}1&s\\0&1\end{array}\right)
\left(\begin{array}{cc}1&0\\-1/f&1\end{array}\right)\\
&=&
\left(
\begin{array}{cc}
 \frac{f^2-f s-s^2}{f^2} & \frac{s (2 f+s)}{f} \\
 -\frac{s}{f^2} & \frac{f+s}{f}
\end{array}
\right)
\end{eqnarray}
The focusing power of the FODO structure is
\begin{equation}
\frac{1}{f_\mathrm{FODO}}=\frac{R_{21}}{R_{11}}=\frac{(1/f)^2 s}{1-s/f-(s/f)^2}
\end{equation}
If the focal length is much larger than the length of the drift region, 
$f \gg s$, the expression simplifies to
\begin{equation}\label{eq-FODO-square}
\frac{1}{f_\mathrm{FODO}}\approx\left(\frac{1}{f}\right)^2 s,
\end{equation}
so in the thin and weak lens approximation, the focusing power of the FODO structure is proportional to the square of the constituent lens' focusing power. This result will be recalled in Section~\ref{sect-pf}.

\section{Electron transfer analysis for grating-type dielectric laser accelerators}\label{sect-tf}

\begin{figure}
\centering
\includegraphics{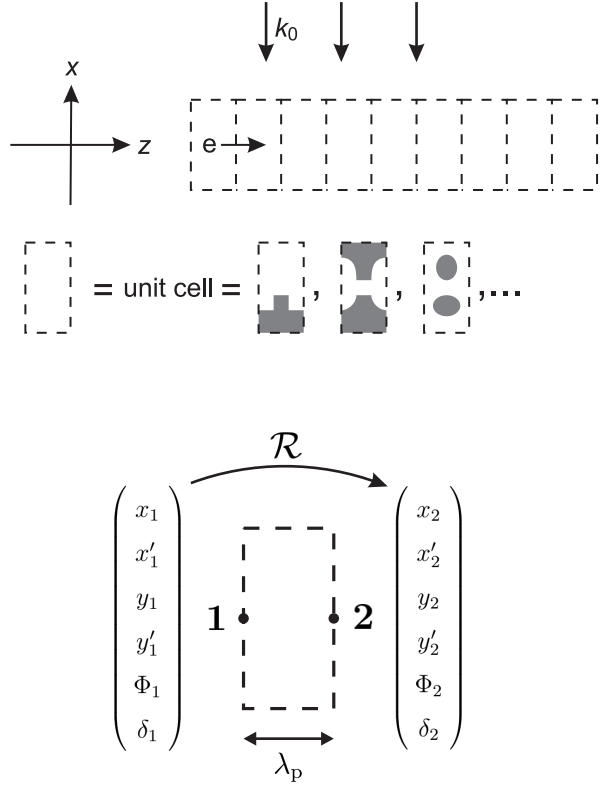}%
\caption%
{%
 \label{fig-gtdla}%
 A segment of a grating-type DLA. The unit cell is iterated in one dimension along the electron beam.
}
\end{figure}

Let us try to develop a methodology, similar to the one outlined in Sect.~\ref{sect-tm}, to describe electron transfer through a grating-type DLA. In this context it is natural to use a Cartesian coordinate system, see Fig.~\ref{fig-gtdla}.
The structure is driven by laser pulses from the direction perpendicular to the electron beam.
It is assumed here that the structure exhibits no large-scale resonances such as guided-mode resonances \cite{Szczepkowicz2016}, so that the filling times are shorter than the laser pulse length. With this assumption
stationary (time-harmonic) calculation of the electromagnetic field is appropriate, and one can obtain realistic time-dependent field by multiplying the stationary result by the laser pulse envelope. This scaling of the result is not carried out here, as it would not affect the conclusions.
Let $(\tilde E_x, \tilde E_y, \tilde E_z, \tilde B_x, \tilde B_y, \tilde B_z)$ represent the stationary solution of the electromagnetic field in a given structure; $E_x(x,y,z,t)=\Re[\tilde E_x(x,y,z) e^{i \omega_0 t}]$ etc.
For a start, assume that electron velocity is perfectly aligned with $\hat z$. If the velocity $\beta_0 c$ is tuned perfectly to the grating period $\lambda_\text p$ and laser wavelength $\lambda_0$,
then $\beta_0=\lambda_\text p/\lambda_0=k_0/k_\text p$, assuming that the DLA is operated at first spatial harmonic \cite{BreuerMcNeurHommelhoff2014}. Let us call $\beta_0c$ the \emph{reference velocity}, corresponding to the
\emph{reference momentum} 
\begin{equation}
p_0=\frac{m\beta_0c}{\sqrt{1-\beta_0^2}},
\end{equation}
Let $\delta$ denote electron's relative deviation from the reference momentum:
\begin{equation}\label{eq-delta}
\delta=\frac{p-p_0}{p_0}.
\end{equation}
Electron position in the transverse plane is described by $(x,y)$, and the slope of the trajectory
is described by 
\begin{equation}
\left(x',y'\right)=
\left(\frac{dx}{dz},\frac{dy}{dz}\right).
\end{equation}

In a radio-frequency accelerator, particle bunch duration $\tau$ is $\sim3$ orders of magnitude smaller 
than the period of the driving electromagnetic wave: $\tau\ll T_0\approx10^{-10}$~s. In contrast to this, in DLA,
the inequality is reversed: $\tau\gg T_0\approx10^{-14}$~s, due to limitations of the present day electron sources
(see eg.~\cite{HoffroggeStein2014}); another limiting factor is the space charge force \cite{BreuerMcNeurHommelhoff2014}. 
As a result, in DLA electrons in a bunch populate all phases.
In the context of grating-type DLAs, phase appears more important than longitudinal position of the electron
along the grating, so it will be convenient to use a parameter $\Phi$ (radians) instead of $S$ (meters) to
describe electron's longitudinal degree of freedom. Let us define $\Phi_1$ of an electron
as the phase of the electromagnetic field at the moment $t_1$ when the electron enters the unit cell of the grating:
\begin{equation}
\Phi_1=\omega_0t_1.
\end{equation}
For an electron with $x'=0,y'=0$ and reference momentum $p=p_0$, traversing the unit cell from $z=z_1$ to $z=z_1+\lambda_p=z_2$,
the phase increases from $\Phi_1$ to $\Phi_1+2\pi=\Phi_2$. Note that in contrast to $x$, $x'$, $y$, $y'$ and $\delta$, the parameter $\Phi$ is not small; it is analogous to the parameter $S$ defined in Sect.~\ref{sect-tm}, not the small parameter $s$. A parameter analogous to $s$ would be $\phi=\Phi_2-\Phi_1-2\pi$.

The set of parameters $(x_1,x'_1,y_1,y'_1,\Phi_1,\delta_1)$ fully describes the classical motion state of a particle
at the entrance of the unit cell. Therefore there exists a \emph{transfer function} $\cal R$, such that
\begin{equation}\label{eq-XRX}
X_2={\cal R} X_1,
\end{equation}
where $X_1=(x_1,x'_1,y_1,y'_1,\Phi_1,\delta_1)^{\mathrm T}$ are the parameters of  the electron at the entrance of the unit cell, and
$X_2=(x_2,x'_2,y_2,y'_2,\Phi_2,\delta_2)^{\mathrm T}$ are the parameters of the electron at the exit of the unit cell, see Fig.~\ref{fig-gtdla}. A matrix-like notation is used here, where one-column matrix $X_2$ is the result of
operator ${\cal R}$ acting on one-column matrix $X_1$.

Using (\ref{eq-XRX}), the properties of $\cal R$ can be studied numerically (particle tracing) even without explicit
formulas for $\cal R$, by specifying sets of example parameters $\{X_1\}$ and calculating corresponding sets of $\{X_2\}$.
Explicit formulas for $\cal R$ are given in Appendix~\ref{sect-equations}; these formulas were used in subsequent analysis.

\section{Example transfer function analysis: a double column structure}\label{sect-lc}

\begin{figure}
\centering
\includegraphics{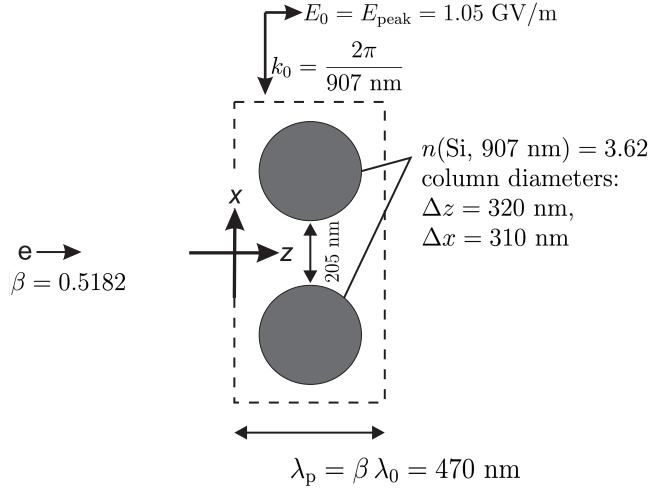}%
\caption%
{%
 \label{fig-lc-geometry}%
 Example parameters used for calculation, based on Ref.~\cite{LeedleCeballos2015}.
}
\end{figure}

\begin{figure}
\centering
\includegraphics{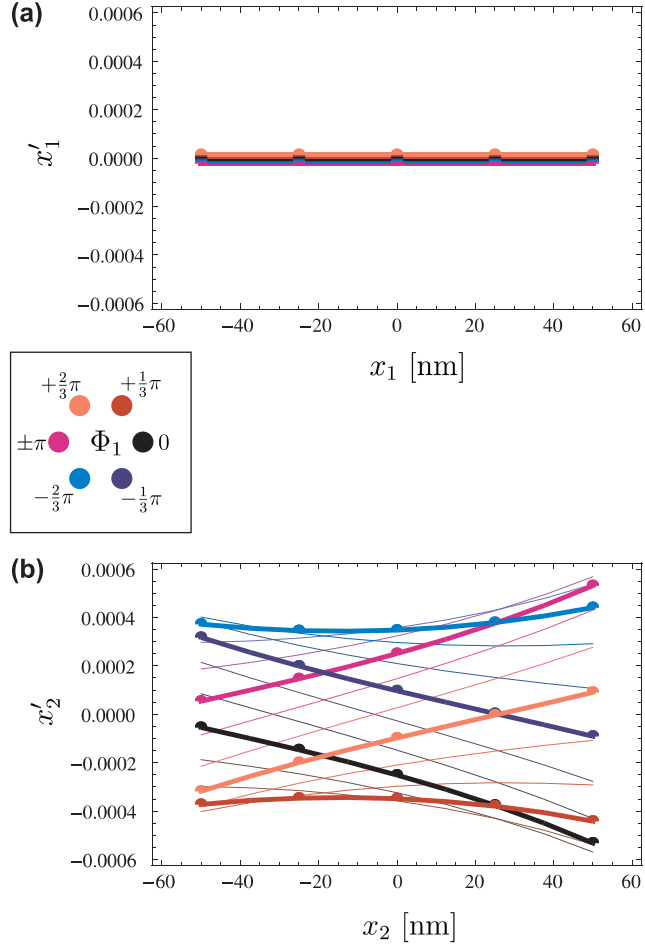}%
\caption%
{%
 \label{fig-lc1}%
 The properties of the transfer function for the unit cell shown in Fig.~\ref{fig-lc-geometry}.
 (a) Assumed $(x_1,x'_1)$ at the entrance of the cell.
 (b) Calculated $(x_2,x'_2)$ at the exit of the cell. 
 The result depends on the initial phase $\Phi_1$ (color coded).
 The thicker lines correspond to six selected phases: $-\pi, -\frac{2}{3}\pi, -\frac{1}{3}\pi, 0, \frac{1}{3}\pi, \frac{2}{3}\pi$.
}
\end{figure}

Let us now apply the concepts of Sect.~\ref{sect-tf} to a specific example of
a grating-type DLA: the double-column structure described in Ref.~\cite{LeedleCeballos2015}.
Figure~\ref{fig-lc-geometry} shows the unit cell. 
The columns are long enough
so that the system can be described in two dimensions $(z,x)$, assuming
infinite column extension in the $\hat y$ direction \cite{LeedleCeballos2015}.
The $y$ coordinate is not significant and will be set to 0.
Let us study some of the properties of the transfer function of the unit cell.
First, the electromagnetic field is calculated using finite element method.
Then the transfer function is applied to sample input parameters
using equations given in Appendix~\ref{sect-equations}. Suppose the incoming electrons are parallel to the $\hat z$ direction: $x'_1=0, y'_1=0$,
and have reference momentum: $p=p_0$, so that $\delta_1=0$. For a start let's choose an initial phase
$\Phi_1=0$ and a set of initial electron positions: $\{x_{1,i}\}=\{-50~\mathrm{nm},-25~\mathrm{nm},0~\mathrm{nm},+25~\mathrm{nm},+50~\mathrm{nm}\}$.
The result of applying $\cal R$ to $X_{1,i}=(x_{1,i},0,0,0,\Phi_1,0)^{\mathrm T}$ is
$X_{2,i}=(x_{2,i},x'_{2,i},0,0,\Phi_{2,i},\delta_{2,i})^{\mathrm T}$. With this
set of calculated parameters various plots are possible. An example is shown
in Fig.~\ref{fig-lc1}, where in (a) pairs $(x_1,x'_1)$ are plotted,
while (b) shows $(x_2,x'_2)$ pairs (black curve).
Subsequently, another initial phase $\Phi_1$ is selected and the procedure is repeated,
with results plotted in different color in the same Figure.

The main question that motivated the described investigations was: is transfer matrix description 
suitable for grating-type DLA structures? The answer follows easily from Fig.~\ref{fig-lc1}.
The transfer function does not in general transform $(0,0,0,0,0,0)^{\mathrm T}$ into
$(0,0,0,0,0,0)^{\mathrm T}$, so it is not a linear function and it cannot be described by a matrix. 
Even if $\Phi$ is
excluded from the set of transformed parameters and one looks for a reduced $\cal R'$
operating in the $(x,x')$ space, Fig.~\ref{fig-lc1} shows that in general
${\cal R'}(0,0)^{\mathrm T}\neq(0,0)^{\mathrm T}$, so matrix description is not possible.
For example, an electron entering the unit cell with phase $\Phi_1=-\frac{2}{3}\pi$ and zero slope
leaves the cell with nonzero slope $x'_2\approx0.0004$.
What is more, neither $\cal R$ nor $\cal R'$ belong to the wider class of affine transforms (linearity with an offset),
because the plots in Fig.~\ref{fig-lc1}(b) are not rectilinear.
Here and in subsequent considerations chromatic effects are neglected: $\delta_1=0$ is assumed.

\begin{figure}
\centering
\includegraphics{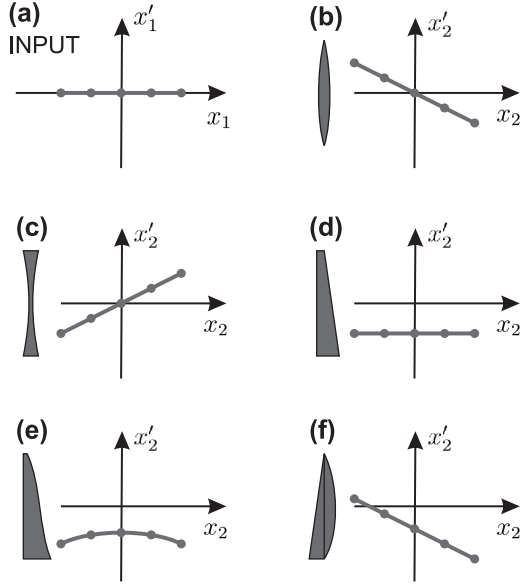}%
\caption%
{%
 \label{fig-glass}%
 The transfer properties of simple optical elements:
 (a) The assumed set of input pairs $(x_1,x'_1)$, representing incident rays with various positions and zero slopes.
 (b)--(f) Calculated $(x_2,x'_2)$ at the exit of (b) converging lens, (c) diverging lens,
 (d) linear prism, (e) nonlinear prism, (f) converging lens with off-axis focus.
 Linear transfer functions describe (a)--(c), for (d), (f) an affine function is needed, while
 for (e) a nonlinear transfer function must be used.
}
\end{figure}

Let us compare the calculated transfer properties with optical transfer 
properties of glass solids, Fig.~\ref{fig-glass}. 
As can be seen form comparison of Figs.~\ref{fig-lc1} and \ref{fig-glass}, the accelerator unit cell,
depending on the incoming electron's phase $\Phi_1$, acts as a 
converging lens for $\Phi_1\in(-\frac{1}{3}\pi,0)$,
a diverging lens for the opposite phase $\Phi_1\in(+\frac{2}{3}\pi,+\pi)$,
an upward-deflecting nonlinear prism (larger deflection for larger $|x_1|$) for $\Phi_1\approx-\frac{2}{3}\pi$,
and a downward-deflecting nonlinear prism for $\Phi_1\approx+\frac{1}{3}\pi$.

\section{Ponderomotive focusing in grating-type dielectric laser accelerators}\label{sect-pf}

In conventional accelerators the primary method of focusing is 
\emph{alternating gradient focusing} (also called \emph{strong focusing}), where lensing quadrupole magnets generate field gradient $\partial B/\partial x, \partial B/\partial y$, and are arranged along the beam direction $z$ with alternating polarity. This is an implementation of the FODO focusing principle described in Section~\ref{sect-tm}. Alternating gradient focusing will be used in planned hybrid accelerator experiments, where a RF beamline will be matched to grating-type DLAs \cite{OdyMusumeci2017, PratBettoni2017}. Of course, the ultimate goal is to develop compact accelerators employing optical-frequency focusing. At present, laser focusing is in early development stage, with conceptual and simulation work under way \cite{PlettnerByer2009, SoongByer2012, WoottonCesar2017}, and a first proof-of-principle experiment with parabolic grating \cite{McNeurKozak2016}. One major problem with focusing in DLA is the same as with acceleration: as yet the phase of electrons in not controlled experimentally, and a shift of phase by $\pi$ reverses the force of the electromagnetic field on the particle and turns focusing into defocusing, so only a fraction of electrons is focused. Is it possible to focus electrons with different phases $\Phi$ at the same time? 

\begin{figure}
\centering
\includegraphics{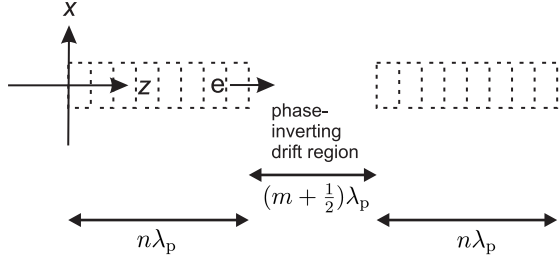}%
\caption%
{%
 \label{fig-gtdla-pf}%
 A DLA accelerator segment analogous to a FODO structure.
}
\end{figure}

An interesting property of a FODO structure is that it keeps its focusing properties if the forces are reversed: both $\cal ODOF$ and $\cal OFOD$ are focusing transformations. Suppose an electron enters a DLA structure shown in Fig.~\ref{fig-gtdla-pf}, and the unit cell has similar
transfer properties as in Fig.~\ref{fig-lc1}.
The transfer function of the whole structure is
\begin{equation}
{\cal R}_\text{tot}={\cal O}_{m+1/2}{\cal R}^n{\cal O}_{m+1/2}{\cal R}^n,
\end{equation}
where again matrix-like notation is used, with multiplication representing mathematical composition of functions, ${\cal R}^n$ denoting the composition of $n$ single cell transfer functions $\cal R$, and ${\cal O}_{m+1/2}$ denoting the linear drift operator (\ref{eq-matrixO}) for $s=(m+\frac{1}{2})\lambda_\text p$.
If, for an electron with phase $\Phi$, $\cal R$  has focusing properties, then ${\cal R}^n$ is also focusing
(for $n$ small enough so that dephasing \cite{McNeurKozak2016} is not significant).
The drift section ${\cal O}_{m+1/2}$ advances the electron phase by $2\pi m+\pi$, so in the second ${\cal R}^n$ section
is defocusing---just like in a FODO structure.
If another electron enters the same structure with phase
$\Phi+\pi$, the structure acts on it as DOFO. For both electrons the structure acts as a converging lens. 
Consider now an electron with such phase $\Phi'$ that
the unit cell acts as a nonlinear upward-deflecting prism.
Now the whole structure cannot be classified as FODO.
After traversing the first ${\cal R}^n$ section,
the electron is deflected upwards, ${\cal O}_{m+1/2}$
reverses the phase, and in the second ${\cal R}^n$ section
the electron is deflected downwards. However, because
the ``prism'' ${\cal R}^n$ is nonlinear, its action is stronger away from the $x=0$ line and the overall
effect of ${\cal R}_\text{tot}$ is again a converging lens.
A similar argument applies to an electron entering
the structure with $\Phi'+\pi$ phase. 
This reasoning, based on $(x,x')$ plots, is purely geometric, but a chromatic effect ($\delta\neq0$) also plays a role in focusing,
as shown in Appendix~\ref{sect-delta}.

\begin{figure}
\centering
\includegraphics{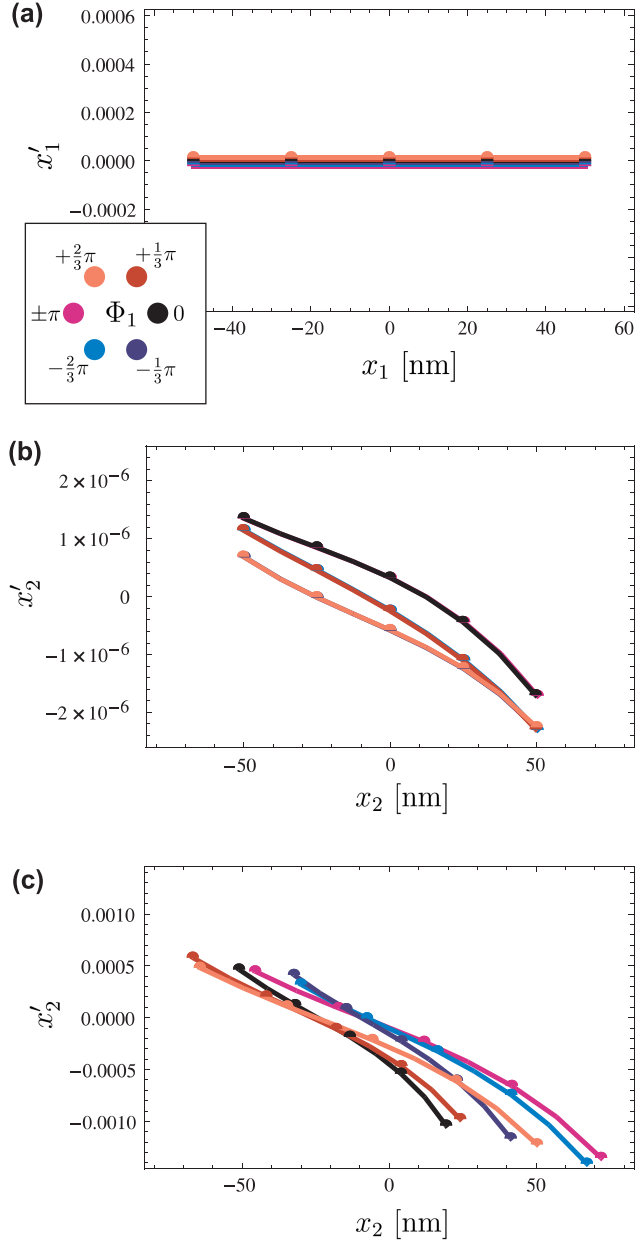}%
\caption%
{%
 \label{fig-pf1}%
 The focusing properties of the accelerator segment shown in Fig.~\ref{fig-gtdla-pf}, calculated for the unit cell shown in
 Fig.~\ref{fig-lc-geometry} using Eq.~(\ref{eq-tf}).
 (a) Assumed $(x_1,x'_1)$ at the entrance of the segment: a parallel beam.
 (b) Calculated $(x_2,x'_2)$ at the exit of the segment, for $n=1$ and $m=0$.
 (c) Calculated $(x_2,x'_2)$ at the exit of the segment, for $n=8$ and $m=5$.
}
\end{figure}

\begin{figure}
\centering
\includegraphics{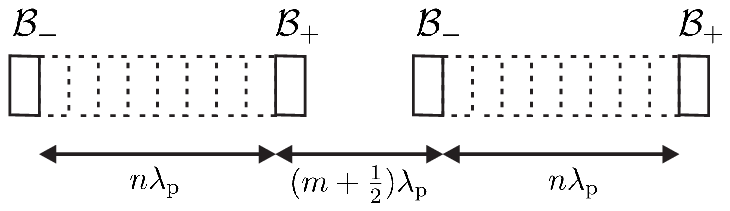}%
\caption%
{%
 \label{fig-gtdla-pf-boundary}%
 Boundary field effects can be handled using boundary cells with corresponding ${\cal B}_\pm$ transfer functions.
}
\end{figure}

The phase-independent focusing effect of ${\cal R}_\text{tot}$ is shown in Fig.~\ref{fig-pf1}.
This structure is a converging lens that exhibits both geometric and ``phase'' aberrations.
The focal lengths for the structure 
${\cal O}_{1/2}{\cal R}{\cal O}_{1/2}{\cal R}$, as shown in Fig.~\ref{fig-pf1}(b), lie
in the range 30~mm--35~mm, so the focusing effect is very weak.
The focal lengths for the structure 
${\cal O}_{5+1/2}{\cal R}^8{\cal O}_{5+1/2}{\cal R}^8$, as shown in Fig.~\ref{fig-pf1}(c), lie
in the range 48~$\mu$m--70~$\mu$m, so the focusing effect
is three orders of magnitude stronger. This shows that grouping of the unit cells
is critical (see also Ref.~\cite{NaranjoValloni2012}). The effect of grouping is even stronger than for a thin lens FODO
structure described by Eq.~(\ref{eq-FODO-square}) (see also Appendix~\ref{sect-ponderomotive-quantitative}). However, grouping
increases the chance of electron collision with the dielectric. It is likely 
that the geometry of the unit cell (Fig.~\ref{fig-lc-geometry}) could be optimized for better transfer and focusing performance,
but this is left for future work. Also, in the presented approach boundary
field effects were neglected. This is justified for large structures like ${\cal O}_{5+1/2}{\cal R}^8{\cal O}_{5+1/2}{\cal R}^8$,
but the calculation of ${\cal O}_{1/2}{\cal R}{\cal O}_{1/2}{\cal R}$ may be inaccurate.
Boundary field effects can be handled with the transfer function approach
by introducing intermediate boundary cells ${\cal B}_\pm$, as shown in Fig.~\ref{fig-gtdla-pf-boundary}.
In this approach, the transfer function ${\cal O}_{5+1/2}{\cal R}^8{\cal O}_{5+1/2}{\cal R}^8$
is an approximation of the more accurate ${\cal O}_{3+1/2}{\cal B}_+{\cal R}^8{\cal B}_-{\cal O}_{3+1/2}{\cal B}_+{\cal R}^8{\cal B}_-$.

\begin{figure}
\centering
\includegraphics{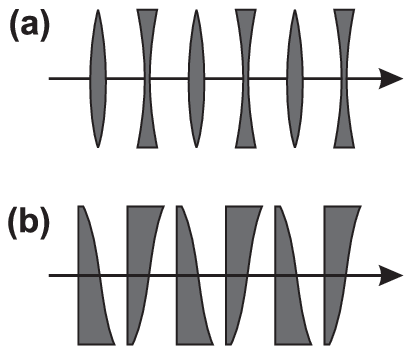}%
\caption%
{%
 \label{fig-lens-and-prisms}%
 Alternating lens (FODO) (a), and alternating nonlinear prisms (b), are examples of ponderomotive focusing systems.
}
\end{figure}

The structure shown in Fig.~\ref{fig-gtdla-pf}, with its converging property, cannot in general (for arbitrary $\Phi$)
be classified as FODO (see Fig.~\ref{fig-lens-and-prisms}), but along with FODO it belongs to a wider class of focusing setups based on
\emph{ponderomotive force} \cite{Mulser1990, Macchi2013} (\cite{Mulser1990} gives historical references). 
Quantitative similarities and differences between the classical ponderomotive force and focusing force of the 
${\cal O}_{m+1/2}{\cal R}^n{\cal O}_{m+1/2}{\cal R}^n$
structure are discussed in Appendix~\ref{sect-ponderomotive-quantitative}.
As noted by Hartman and Rosenzweig \cite{HartmanRosenzweig1993}, other alternating focusing schemes used in radio frequency accelerators, like radio-frequency quadrupole (RFQ) focusing \cite{Wangler2008}, or alternating phase focusing \cite{Wangler2008,Swenson1976}, are also based on ponderomotive force.
In the context of DLA, a ponderomotive focusing scheme has already been studied for photonic band-gap accelerators \cite{NaranjoValloni2012}. For grating-type DLA, the idea was considered in Ref.~\cite{BreuerMcNeurHommelhoff2014}
(citing \cite{Swenson1976,NaranjoValloni2012}), but specific implementation was not proposed.

Ponderomotive focusing of electrons in the transverse plane is analogous to the redistribution of sand on a Chladni plate \cite{Chladni1787}. A grain of sand on a vibrating plate is subject to alternating force whose amplitude
is a function of position on the plate, and diffuses towards regions of smaller amplitude, finally settling
in the nodal regions.
Similarly, an electron traversing a ${\cal O}_{m+1/2}{\cal R}^n{\cal O}_{m+1/2}{\cal R}^n{\cal O}_{m+1/2}{\cal R}^n\ldots$
structure with reference velocity $\beta_0 c$ is subject to an alternating force of frequency
\begin{equation}
\label{eq-pffreq}
\omega'=\frac{\omega_0}{(2n+2m+1)}
\end{equation}
(because the spatial period of the structure is $(2n+2m+1)\lambda_\text p$),
and is attracted in the transverse plane towards regions of smaller force amplitude---smaller electromagnetic field.
The field is stronger close to the dielectric surfaces, and for double grating-like structures the minimum
of the transverse force lies in the electron channel between the two surfaces.

\section{Conclusion and outlook}

Transfer matrices are known to be useful for the description of particle motion through the segments of conventional RF accelerators. 
A similar description is proposed here for grating-type DLAs: 
linear transfer functions, represented by matrices,
are replaced by nonlinear transfer functions;
matrix multiplication is replaced by numerical function composition; 
these differences are hardly noticeable with a compact, matrix-like notation.
The approach facilitates quantitative 
description of electron transfer through a DLA structure, where, in the first approximation,
the transfer properties of larger units are easily determined from the transfer properties of the 
DLA unit cell.
Hopefully this approach will make easier the conceptual and simulation work on new designs, 
and help in clear presentation and discussion of the properties of new DLA structures.
One example of presentation of transfer properties are the $(x,x')$ plots, 
sometimes called ``trace space plots'' in the RF accelerator literature;
such plots are already entering the DLA literature \cite{OdyMusumeci2017}, 
and can naturally be produced with the transfer function approach described here.

In Sect.~\ref{sect-pf} the transfer function approach led naturally to the idea of building
a FODO-like DLA structure, which focuses electrons irrespective of the phase.
The converging force in the proposed setup is yet another example of ponderomotive force.
Further work is required to optimize the geometry. One approach would be to drive the 
structure symmetrically from two sides by employing distributed Bragg reflectors \cite{PratBettoni2017}.

In this paper the transfer function is applied only to lensing properties of DLA structures.
Of course the primary function of DLAs is to \emph{accelerate}: to increase
$\delta$. Here it was assumed that $\delta_1=0$ and $\delta_2$ was not analyzed.
Hopefully the described formalism with its six parameters $(x,x',y,y',\Phi,\delta)$ 
will also be useful to describe acceleration schemes.
Here a major challenge is the phase distribution of electrons, which results in
only a fraction of electrons being accelerated. To address this issue,
methods to compress the particle bunch are investigated \cite{PratBettoni2017} to obtain single-phase particles.
More generally, a method is needed to redistribute the electron phases to populate several narrow 
$\Phi$ subsets separated by $2\pi$. Alternatively, perhaps an accelerating
scheme working for all incoming $\Phi$ could be invented.
Formulation of these challenges using $(x,x',y,y',\Phi,\delta)$ may
accelerate progress in this field.

\begin{acknowledgements}

I am grateful to Martin Koz{\'a}k, Joshua McNeur and Peter Hommelhoff for inspiring discussions during my visit in Erlangen in May 2016.

I am grateful to Wroc{\l}aw Networking and Supercomputing Center for granting access to the PLATON computing infrastructure.

\end{acknowledgements}


\appendix
\section{Transfer function equations}\label{sect-equations}

The transfer function defined by Eq.~(\ref{eq-XRX}) can put into 
the following explicit form (derived in Appendix~\ref{sect-derivation}):
\begin{subequations}\label{eq-tf}
\begin{eqnarray}
x_2&=&x_1+x'_1(z_2-z_1))\label{eq-tf-x}\\[1mm]
x'_2&=&\frac
{x'_1+\frac{\Delta p_x}{C p_0(1+\delta_1)}}%
{1+\frac{\Delta p_z}{C p_0(1+\delta_1)}}\label{eq-tf-xp}\\[1mm]
y_2&=&y_1+y'_1(z_2-z_1))\label{eq-tf-y}\\[1mm]
y'_2&=&\frac
{y'_1+\frac{\Delta p_y}{C p_0(1+\delta_1)}}%
{1+\frac{\Delta p_z}{C p_0(1+\delta_1)}}\label{eq-tf-yp}\\[1mm]
\Phi_2&=&\Phi_1+k_0\frac{z_2-z_1}{\beta_z}\label{eq-tf-Phi}\\[1mm]
\delta_2&=&{\textstyle
(1+\delta_1)C\sqrt{
\left(x'_1{+}\frac{\Delta p_x}{C p_0 (1+\delta_1)}\right)^{\!2}{\!\!+\!}
\left(y'_1{+}\frac{\Delta p_y}{C p_0 (1+\delta_1)}\right)^{\!2}{\!\!+\!}
\left(1   {+}\frac{\Delta p_z}{C p_0 (1+\delta_1)}\right)^{\!2}
}
-1
}\label{eq-tf-delta}
\end{eqnarray}
\end{subequations}
In Eqations~(\ref{eq-tf}), the following auxiliary quantities were used:
$C$ is the trajectory deflection cosine = $\hat z\cdot \hat v_1$,
$\beta_z$ is the relative longitudinal velocity,
$\Delta p_x, \Delta p_y, \Delta p_z$ is the momentum change of the electron.
The formulas for these auxiliary quantities are:
\begin{subequations}
\label{eq-tf-aux}
\begin{eqnarray}
C&=&\frac{1}{\sqrt{{x'_1}^2+{y'_1}^2+1}}\label{eq-tf-aux-C}
\\
\beta_z&=&C\frac{p_0(1+\delta_1)}{\sqrt{p_0^2(1+\delta_1)^2+m^2c^2}}\label{eq-tf-aux-betaz}
\\
\Delta p_x&=&\Re\left\{
\frac{(-e)}{c}
\int_{z_1}^{z_2}
\left(\frac{1}{\beta_z}\tilde E_x+y'_1c\tilde B_z-c\tilde B_y \right)
\exp\left[i\left(\Phi_1+k_0\frac{z-z_1}{\beta_z}\right)\right]
dz
\right\}\label{eq-tf-aux-dpx}
\\
\Delta p_y&=&\Re\left\{
\frac{(-e)}{c}
\int_{z_1}^{z_2}
\left(\frac{1}{\beta_z}\tilde E_y+c\tilde B_x-x'_1c\tilde B_z \right)
\exp\left[i\left(\Phi_1+k_0\frac{z-z_1}{\beta_z}\right)\right]
dz
\right\}\label{eq-tf-aux-dpy}
\\
\Delta p_z&=&\Re\left\{
\frac{(-e)}{c}
\int_{z_1}^{z_2}
\left(\frac{1}{\beta_z}\tilde E_z+x'_1c\tilde B_y-y'_1c\tilde B_x \right)
\exp\left[i\left(\Phi_1+k_0\frac{z-z_1}{\beta_z}\right)\right]
dz
\right\}\label{eq-tf-aux-dpz}
\end{eqnarray}
\end{subequations}
The complex-valued functions $(\tilde E_x, \tilde E_y, \tilde E_z, \tilde B_x, \tilde B_y, \tilde B_z)$ represent the stationary solution of the electromagnetic field in a given structure; $E_x(x,y,z,t)=\Re[\tilde E_x(x,y,z) e^{i \omega_0 t}]$ etc.
The components of the electromagnetic field under the integrals are taken at the electron position parameterized by $z$:
\begin{equation}\label{eq-xyz}
(x,y,z)=(x_1+x'_1(z-z_1), y_1+y'_1(z-z_1), z).
\end{equation}
It is assumed here that the motion of the electrons is piecewise linear,
with straight line trajectory within one unit cell, from $z_1$ to $z_2=z_1+\lambda_p$;
although the electron accumulates momentum
during its flight through the cell, in calculation the accumulated momentum
is added only at the exit of the cell; this is equivalent to the Euler method
of solving differential equations (a first-order Runge-Kutta method). This method
is numerically less efficient than the conventional fourth-order Runge-Kutta algorithm,
but the formulas are simpler, easier to derive, analyze, expand in series, and this facilitates elementary physical insight.

The validity of the Euler approximation was checked for the calculations of Sect.~\ref{sect-lc} and \ref{sect-pf} by subdividing the unit cell into 4 sub-cells, and calculating the unit cell transfer function as a composition 
${\cal R}={\cal R}_{4}{\cal R}_{3}{\cal R}_{2}{\cal R}_{1}$, 
where ${\cal R}_{1}$ is the transfer function from $z$ to $z+\frac{1}{4}\lambda_\text p$, etc. Such refinement did not influence the $(x_2,x'_2)$ plots in Figs.~\ref{fig-lc1} and \ref{fig-pf1}(c). On the other hand, the refinement did quantitatively influence the calculation shown in
Fig.~\ref{fig-pf1}(b), where the accelerator segment consisted of only two elementary cells.
In this case the calculation converged for $n\approx50$ subdivision segments, and this large number
of segments was used to produce Fig.~\ref{fig-pf1}(b).

Equations~(\ref{eq-tf}) contain small dimensionless parameters $x'$, $y'$, $\delta$, $\Delta p_i/p_0$.
In textbooks on conventional accelerators such equations are usually
expanded in Taylor series and higher order terms are dropped \cite{Wille2000}.
For the purposes of this paper Taylor expansion of Eq.~(\ref{eq-tf}) would not be productive.
Note that linearization of the transfer function is not possible, as discussed in Sect.~\ref{sect-lc}.

\section{Derivation of the transfer function equations}\label{sect-derivation}

Assuming the electron trajectory is linear within the unit cell (or its subset, see previous section),
as the electron travels from $z_1$ to $z_2$, its transverse position $x$ increases from
$x_1$ to $x_2$, with $x_2=x_1+\Delta x=x_1+ \frac{\Delta x}{\Delta z}\Delta z=x_1+x'_1(z_2-z_1)$.
Similarly, $y_2=y_1+y'_1(z_2-z_1)$. 

$C$ is the cosine of the deflection of electron trajectory from the $\hat z$ direction,
$C=\hat z\cdot\hat v_1=(0,0,1)\cdot\frac{(v_{1x},v_{1y},v_{1z})}{v_1}$, where $\vec v_1$ is the velocity
of the electron at the entrance of the cell. It follows that
$C=\frac{v_{1z}}{v_1}=\frac{v_{1z}dt}{v_1 dt}=\frac{dz}{\sqrt{dx^2+dy^2+dz^2}}
=\frac{1}{\sqrt{(dx/dz)^2+(dy/dz)^2+(dz/dz)^2}}=\frac{1}{\sqrt{{x'_1}^2+{y'_1}^2+1}}$.

Momentum and velocity at the entrance of the cell are related by
$p_1=\frac{1}{\sqrt{1-{\beta_1}^2}}m\beta_1 c$, or $\beta_1=\frac{p_1}{\sqrt{{p_1}^2+m^2c^2}}$, or, using the definition of
$\delta$ (Eq.~\ref{eq-delta}), $\beta_1=\frac{p_0(1+\delta_1)}{\sqrt{{p_0}^2(1+\delta_1)^2+m^2c^2}}$.
The $z$ component of the relative velocity is $\beta_{1z}=\frac{\beta_{1z}}{\beta_1}\beta_1=C\frac{p_0(1+\delta_1)}{\sqrt{{p_0}^2(1+\delta_1)^2+m^2c^2}}$.

The slope at the exit of the cell is $x'_2=\frac{dx}{dz}=\frac{p_{2x}}{p_{2z}}=\frac{p_{1x}+\Delta p_x}{p_{1z}+\Delta p_z}=
\frac{{p_{1x}}/{p_{1z}}+{\Delta p_x}/{p_{1z}}}{{p_{1z}}/{p_{1z}}+{\Delta p_z}/{p_{1z}}}=
\frac{x'_1+{\Delta p_x}/{p_{1z}}}{1+{\Delta p_z}/{p_{1z}}}=
\frac{x'_1+{\Delta p_x}/{C p_1}}{1+{\Delta p_z}/{C p_1}}=
\frac{x'_1+{\Delta p_x}/{C p_0(1+\delta_1)}}{1+{\Delta p_z}/{C p_0(1+\delta_1)}}$.
The expression for $y'_{2}$ is analogous.

The momentum at the exit of the cell is\\
$p_2=\sqrt{(p_{1x}+\Delta p_x)^2+(p_{1y}+\Delta p_y)^2+(p_{1z}+\Delta p_z)^2}$\\
${}=p_{1z}\sqrt{\left(\frac{p_{1x}}{p_{1z}}+\frac{\Delta p_x}{p_{1z}}\right)^2+
\left(\frac{p_{1y}}{p_{1z}}+\frac{\Delta p_y}{p_{1z}}\right)^2+
\left(\frac{p_{1z}}{p_{1z}}+\frac{\Delta p_z}{p_{1z}}\right)^2}$\\
${}=C p_1
\sqrt{\left(x'_1+\frac{\Delta p_x}{C p_1}\right)^2+
\left(y'_1+\frac{\Delta p_y}{C p_1}\right)^2+
\left(1+\frac{\Delta p_z}{C p_1}\right)^2}$\\
${}=C p_0(1+\delta_1)
\sqrt{\left(x'_1+\frac{\Delta p_x}{C p_0(1+\delta_1)}\right)^2+
\left(y'_1+\frac{\Delta p_y}{C p_0(1+\delta_1)}\right)^2+
\left(1+\frac{\Delta p_z}{C p_0(1+\delta_1)}\right)^2}$,\\
so the relative momentum deviation is by definition (\ref{eq-delta}) 
$\delta_2=p_2/p_0-1$\\
${}=C(1+\delta_1)
\sqrt{\left(x'_1+\frac{\Delta p_x}{C p_0(1+\delta_1)}\right)^2+
\left(y'_1+\frac{\Delta p_y}{C p_0(1+\delta_1)}\right)^2+
\left(1+\frac{\Delta p_z}{C p_0(1+\delta_1)}\right)^2}
-1
$

The phase increases from $\Phi_1=\omega_0 t_1$ to $\Phi_2=\omega_0 t_2$,
and $\Phi_2=\Phi_1+\omega_0\Delta t=\Phi_1+\omega_0\frac{\Delta z}{\beta_z c}=\Phi_1+\frac{k_0(z_2-z_1)}{\beta_z}$.

During its flight through the cell the electron receives
momentum $(\Delta p_x,\Delta p_y,\Delta p_z)$ from the electromagnetic field, where
$\Delta p_x=\int F_x dt=\int\frac{F_x}{dz/dt}dz=\int\frac{F_x}{c\beta_z}dz
=\frac{(-e)}{c}\int\frac{1}{\beta_z}(E_x+\beta_y c B_z - \beta_z c B_y)dz
=\frac{(-e)}{c}\int(\frac{1}{\beta_z}E_x+y'_1 c B_z - c B_y)dz$. 
The electromagnetic field components under the integral are taken 
at the electron location, parameterized by $z$:
$E_x=E_x(x,y,z,t)=\Re\{\tilde E_x(x,y,z)\exp[i\omega_0 t]\}
=\Re\{\tilde E_x(x(z),y(z),z)\exp[i\omega_0 t(z)]\}
=\Re\{\tilde E_x(x_1+x'_1(z-z_1),y_1+y'_1(z-z_1),z)\exp[i(\Phi_1+\frac{k_0(z-z_1)}{\beta_x})]\}$,
and similarly for $E_y$ and $E_z$.
The real-part operator $\Re$ is additive and in the expression for
$\Delta p_x$ can act as the final operation:
$\Delta p_x=
\Re\{
\frac{(-e)}{c}\int(\frac{1}{\beta_z}\tilde E_x+y'_1 c \tilde B_z - c \tilde B_y)\exp[i(\Phi_1+\frac{k_0(z-z_1)}{\beta_x})]dz
\}$. 
The derivation of expressions for $\Delta p_y$ and $\Delta p_z$ is similar.

\section{Are variations in $\delta$ significant for focusing?}\label{sect-delta}

\begin{figure}
\centering
\includegraphics{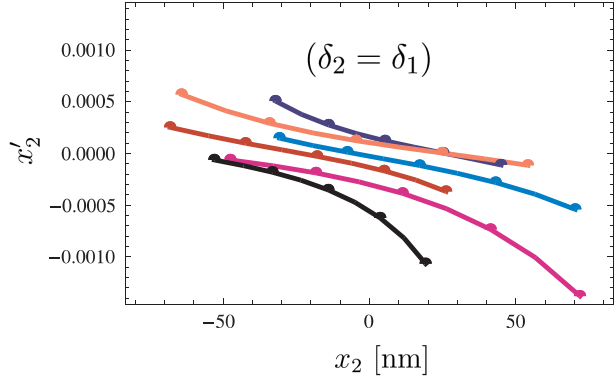}%
\caption%
{%
 \label{fig-pf2}%
 The focusing properties of the ${\cal O}_{5+1/2}{\cal R}^8{\cal O}_{5+1/2}{\cal R}^8$ segment calculated using ,,spoiled'' ($\delta_2=\delta_1$) transfer function (compare with the correct result in Fig.~\ref{fig-pf1}(c)).
}
\end{figure}

In Section \ref{sect-pf} the forces on an electron traversing a ${\cal O}_{m+1/2}{\cal R}^n{\cal O}_{m+1/2}{\cal R}^n$ structure are
discussed, and it is shown that the overall effect is focusing. The argument, based on $(x,x')$ plots for a single cell, 
is purely geometric, assuming $\delta=0$ and thus neglecting the ,,chromatic effects''. However, the calculations
leading to Fig.~\ref{fig-pf1} are exact in the sense that full transfer function is used (equations (\ref{eq-tf})),
so in the calculation $\delta$ is nonzero (except the entrance of the cell). Is focusing modified by chromatic effects ($\delta\neq 0$)?
To answer this question, let us ``spoil'' the transformation (\ref{eq-tf}) by assuming $\delta_2=\delta_1$ instead
of Eq.~(\ref{eq-tf-delta}). This means that now $\delta$ is forced to remain constant, equal to the initial zero value,
and that the phase $\Phi$ advances in each elementary cell by exactly $2\pi$.
The result is shown in Fig.~\ref{fig-pf2}.
The structure still has focusing properties, but the result is significantly
different than for the correct transformation, and the average focusing power decreases by a factor of $\sim2$.
So the ,,geometric argument'', while essentially correct, does not capture all focusing factors,
and chromatic effects are also important.

\section{Ponderomotive focusing and ponderomotive force -- quantitative analysis}\label{sect-ponderomotive-quantitative}

\begin{figure}
\centering
\includegraphics{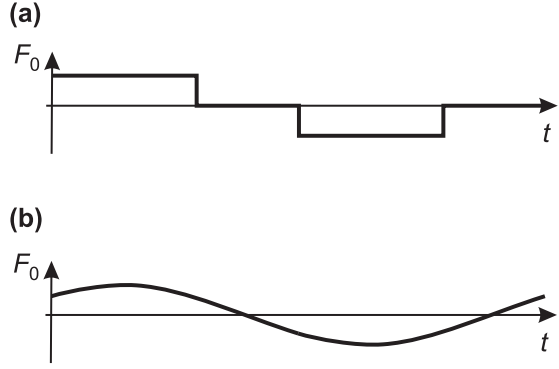}%
\caption%
{%
 \label{fig-oscillation}%
 (a) Time dependence of the focusing force in a FODO-like DLA cell (Fig.~\ref{fig-gtdla-pf}).
 (b) Harmonic oscillation leading to classical ponderomotive force.
}
\end{figure}

Suppose a particle is subject to an oscillating force $\vec F=\vec F_0 \cos \omega t$,
whose amplitude $\vec F_0$ varies spatially on length scales larger than the amplitude
of the $\omega$--oscillation of the particle. Under these circumstances an effective, average
force on the particle arises, called the ponderomotive force (see e.g. \cite{Mulser1990, Macchi2013}):
\begin{equation}
\label{eq-pf}
\vec F_{\text p}\sim - \frac{1}{\omega^2}\nabla(|\vec F_0|^2).
\end{equation}
For a high-energy particle traversing a FODO-like DLA structure described in Sect.~\ref{sect-pf},
the transverse defecting force is a function of transverse position $(x,y)$ and oscillates with frequency
$\omega'$ given by Eq.~(\ref{eq-pffreq}), causing small-amplitude
electron oscillation in the $(x,y)$ plane, so the basic requirements for \emph{ponderomotive force} are satisfied. 
The distinction between a single ,,FODO'' cell and repeated ,,FODOFODO\ldots'' structure does not
affect the physical focusing mechanism and should not affect the terminology.
There is however one significant difference between the classical ponderomotive force and the present
situation: the oscillation of the focusing force is not harmonic, as shown in Fig.~\ref{fig-oscillation}.
This sheds doubt on the applicability of Eq.~(\ref{eq-pf}) to the present situation.
Te derivation of this equation \cite{Mulser1990, Macchi2013} should be reconsidered,
allowing for non-harmonic force oscillations, which is beyond
the scope of this paper. Nevertheless, let us numerically check three
features of ponderomotive focusing occuring in the ${\cal O}_{5+1/2}{\cal R}^8{\cal O}_{5+1/2}{\cal R}^8$ structure, 
and compare them with Eq.~(\ref{eq-pf}).

(1) Let us reduce the amplitude of force oscillation $F_0$ by half by reducing driving laser amplitude $E_0$ (see Fig.~\ref{fig-lc-geometry}) by half. The calculation yields the result that the average focusing power of the structure
decreases by a factor of 4.2, signifying the decrease of the ponderomotive focusing force by
the same factor. This result is close to the value of 4 expected from Eq.~(\ref{eq-pf}).

(2) Let us shorten the structure approximately by half: ${\cal O}_{3+1/2}{\cal R}^4{\cal O}_{3+1/2}{\cal R}^4$.
This increases the oscillation frequency by a factor of 2.
The result is that the average focusing power decreases by a factor of 7.5. This is actually
closer to $2^3$ than to the value $2^2$ expected from Eq.~(\ref{eq-pf}) and questions the applicability of 
this equation to non-harmonic oscillating forces.

\begin{figure}
\centering
\includegraphics{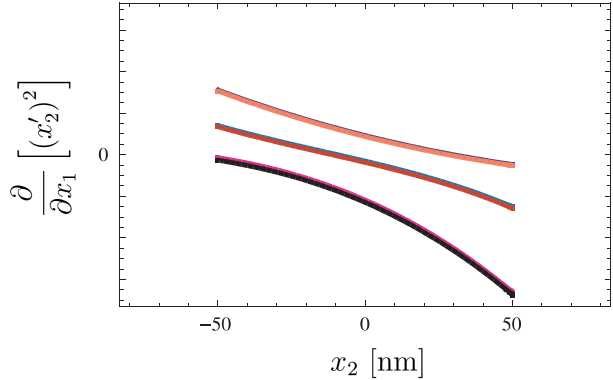}%
\caption%
{%
 \label{fig-final}%
 Equation (\ref{eq-final}), based on the transfer
 properties of an elementary DLA cell, plotted for
 the same six phases as in Fig.~\ref{fig-pf1}.
}
\end{figure}

(3) Let us, for the structure ${\cal O}_{5+1/2}{\cal R}^8{\cal O}_{5+1/2}{\cal R}^8$, calculate the gradient  $\nabla(F_0^2)=\frac{\partial}{\partial x}(F_0^2)$. The force amplitude $F_0$ is equal to the average force
exerted on the electron traversing an elementary DLA cell
\begin{equation}
F_0=\langle F_x \rangle\sim \Delta p_x\sim \Delta x'=x'_2-x'_1,
\end{equation}
where an approximate form $x'_2=x'_1+\Delta p_x/p_0$ of Eq.~(\ref{eq-tf-xp}) was used.
For simplification, a transfer function for $x'_1=0$ is considered here:
\begin{equation}
(x_1,0,0,0,\Phi_1,0)  \xrightarrow{\displaystyle \cal R}  (x_2,x'_2,0,0,\Phi_2,\delta_2),
\end{equation}
so the force amplitude $F_0$ is simply proportional to $x'_2$ for an elementary cell,
where $x'_2$ is a function of six parameters: $x'_2=x'_2(x_1,0,0,0,\Phi_1,0)$.
The final equation for the gradient, neglecting multiplicative constants, is
\begin{equation}
\label{eq-final}
\nabla(F_0^2)\sim \frac{\partial}{\partial x_1}\left[\left( x'_2\right)^2\right].
\end{equation}
This functional dependence is plotted in Fig.~\ref{fig-final}.
If Eq.~(\ref{eq-pf}) was strictly valid,
the plots in Figs~\ref{fig-pf1}(c) and \ref{fig-final} should
be the same up to a multiplicative constant. While
both plots indicate focusing, there are quantitative
differences, so Eq.~(\ref{eq-pf}) is not strictly
valid for DLA ponderomotive focusing force.

There could be one more reason for the inaccuracy of Eq.~(\ref{eq-pf}) in the present
situation. Perhaps the transverse oscillation amplitude of the electron is too large.
This hypothesis mav be verified in future work.





%

\end{document}